# Numerical simulation of liquid film formation and its heat transfer through vapor bubble expansion in a microchannel

Junnosuke Okajima[1], Peter Stephan[2,3]

[1]Institute of Fluid Science, Tohoku University, 2-1-1, Katahira, Aoba-ku, Sendai, Miyagi 980-8577, Japan

[2]Institute of Technical Thermodynamics, Technische Universität Darmstadt, Alarich-Weiss-Straße 10, 64287 Darmstadt, Germany

[3]Center of Smart Interfaces, Technische Universität Darmstadt, Alarich-Weiss-Straße 10, 64287 Darmstadt, Germany

The evaporation of vapor bubbles inside a microchannel is important to realize a device with high cooling performance. The liquid film formed on the solid surface is essential for evaporative heat transfer from solid to fluid; its formation process and heat transfer characteristics need to be investigated. The expansion process of a single vapor bubble via evaporative heat transfer in microchannels was evaluated via a numerical simulation in this study. In the calculation model, the working fluid used was saturated FC-72 at 0.1013 MPa and the channel diameter was 200 µm. The superheat of the initial temperature field and wall were considered as parameters. To evaluate the heat transfer characteristics, the time variation of liquid film thickness was evaluated. The averaged liquid film thickness had a correlation with the capillary number. Additionally, the dominant heat transfer mode was estimated by decomposing the heat transfer rate into the heat-transfer rate through the liquid film, rear edge, and wake. When the superheat was low, the heat transfer mostly occurred via liquid film evaporation; the heat flux through the liquid film could be predicted using the liquid film thickness. On the other hand, in cases of higher superheat, owing to rapid expansion of the vapor bubble, no evaporative heat transfer occurred through the liquid film around the bubble head. It could be inferred from this study that the relationship between the thickness of the thermal boundary layer of the bubble and liquid film thickness is important for predicting the cooling effect of this phenomena. When the vapor bubble grows in the high superheat liquid, the rapid growth makes the liquid film thick, and the thick liquid film prevents the heat transfer between the liquid–vapor interface and heated wall.

Corresponding author:

Junnosuke Okajima

Address: Institute of Fluid Science, Tohoku University, 2-1-1, Katahira, Aoba-ku, Sendai, Miyagi 980-8577, Japan

TEL & FAX: +81-22-217-5879

E-mail: j.okajima@tohoku.ac.jp

## 1. Introduction

During nucleate boiling or two-phase flow with evaporation, a liquid film is formed on the solid wall, and its thickness strongly affects the heat transfer from solid to fluid. The liquid film evaporation offers a high potential for providing cooling effects. Therefore, the thickness of the liquid film and formation process are important factors to consider in a cooling device. To achieve high heat flux cooling, the relationship between the thickness of the liquid film and its heat transfer is required. It was proved in a previous study that the heat flux caused by liquid film evaporation in microchannels exceeds the critical heat flux of nucleate boiling under specific conditions[1].

There are many studies pertaining to the measurement of liquid film thickness. Monde[2] conducted an experiment to estimate the liquid film thickness from the temperature variation on a heated wall. In this experiment, the liquid film was formed by air bubbles. Moriyama et al.[3] evaluated the liquid film thickness of evaporating bubbles in the gap between two parallel plates. They found a correlation among the liquid film thickness, capillary number, and accelerated Bond number. Han et al.[4] measured the liquid film thickness of the air/liquid two-phase flow in a minichannel with a laser focus displacement meter. They modeled their experimental data using the Reynolds number, Weber number, and capillary number. Furthermore, Han et al. investigated the effect of acceleration of the bubble[5] and measured the liquid film thickness in the evaporation process[6]. Utaka et al. developed a laser extinction method to measure the thickness of the liquid film in the narrow gap between two flat plates[7] and discussed the heat transfer characteristics of the liquid film[8]. Scammell and Kim[9] measured the thickness of the liquid film and temperature field in the flow with Taylor bubble and derived the importance of turbulent mixing by bubble wake. However, the basic characteristics of the liquid film during evaporation have not been clarified yet.

Many recent studies focused on performing numerical simulations to understand the phase change phenomena. The numerical simulation of single bubble evaporation in a microchannel has been reported under various conditions: heating with a constant temperature wall[10], heating with a constant heat flux wall[11], focusing on coalescence behavior[12, 13], focusing on the transition from slug to annular flow[14], and considering the thermocapillary effect[15]. Magnini and Thome[16] and Ferrari et al.[17] conducted numerical simulations of slug flow boiling to evaluate the effect of the liquid film on heat transfer under a constant heat flux.

In previous studies, however, owing to the constant heat flux on the heated wall, convective and evaporative heat transfers were mixed and it was difficult to identify the evaporation effect from the mixed heat transfer. Additionally, the contribution of the liquid film around the vapor bubble on the heat transfer between the heated wall and liquid−vapor interface has not been clarified yet. Hence, the process of liquid film formation and heat transfer needs to be investigated. The objectives of this study were to develop a numerical model to calculate the phase change heat transfer in a microchannel and clarify the basic characteristics of the liquid film formed during a single bubble expansion in a microchannel. We specifically focused on the basics of the dynamic behavior of the liquid film; therefore, the isothermal field in the initial condition and a constant temperature wall as the boundary condition were considered to simplify the problem.

## 2. Analysis model

Fig. 1 shows the two-dimensional axisymmetric analysis model used in this study. The radius and length of channel are 100 μm and 5 mm, respectively. The initial bubble that is half the size of the channel is placed at the inlet.

As mentioned earlier, it is important to evaluate only the evaporative heat transfer via the vapor bubble. Therefore, an isothermal field was used in this study. The same superheat *ΔT* was given at the wall and in the initial temperature field. Owing to this treatment, no heat transfer occurred between the heated wall and initial liquid. Only the evaporative heat transfer via the vapor bubble affects the temperature variation at the wall. Additionally, a perfect wet wall was assumed in this study.

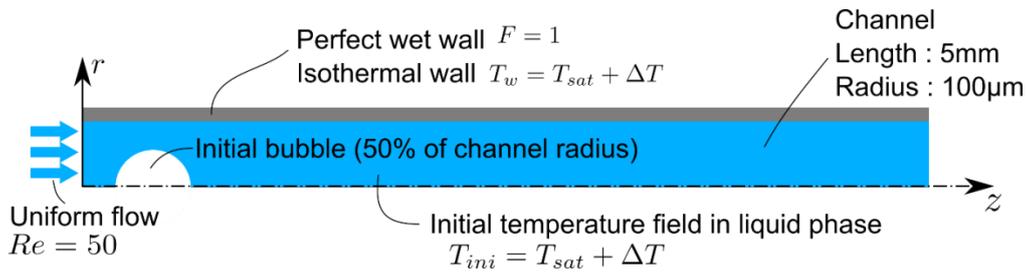

Fig. 1. Analysis model

## 3. Governing equation and numerical method

To solve the phase change heat transfer, the calculation method proposed by Kunkelmann et al.[18] was used in this study. The governing equations are the continuity equation, incompressible Navier–Stokes equation, and energy equation:

$$\nabla \cdot \boldsymbol{u} = \Sigma_V, \qquad (1)$$

$$\rho \left( \frac{\partial \boldsymbol{u}}{\partial t} + (\boldsymbol{u} \cdot \nabla)\boldsymbol{u} \right) = -\nabla p + \nabla(\mu \nabla \boldsymbol{u}) + \sigma \kappa \nabla F, \qquad (2)$$

$$\frac{\partial(\rho c T)}{\partial t} + \nabla \cdot (\rho c \boldsymbol{u} T) = \nabla \cdot (k \nabla T) + \Sigma_e, \qquad (3)$$

where $\boldsymbol{u}$ [m/s] is the velocity vector, $\rho$ [kg/m$^3$] is the density, $p$ [Pa] is the pressure, $\mu$ [Pa·s] is the viscosity, $\sigma$ [N/m] is the surface tension, $\kappa$ [1/m] is the curvature of the interface, $c$ [J/(kg·K)] is the specific heat, $T$ [K] is the temperature, $\Sigma_V$ [1/s] is the source term of volume due to phase change, and $\Sigma_e$ [W/m$^3$] is the energy source term. The surface tension force was modeled using the continuum surface force model developed by Brackbill et al.[19]. The counter-based volume-of-fluid (VOF) method proposed by Kunkelmann et al.[18] was used to capture the interface between the liquid and vapor phase. Here, *F* [-] denotes the volume fraction of the liquid phase, and the liquid fraction *F* was advected according to the advection equation

$$\frac{\partial F}{\partial t} + \nabla \cdot (\boldsymbol{u} F) + \nabla \cdot (c_F |\boldsymbol{u}| \boldsymbol{n}_{int} F(1-F)) = \Sigma_V F, \tag{4}$$

where $c_F$ is the compression parameter, which scales with the flow velocity. In this study, $c_F$ was fixed as unity. From the VOF field, the sharp interface was reconstructed. The evaporative mass flux was calculated using the following equations:

$$\dot{m}_{evp} h_{LV} = q_{evp,L} + q_{evp,V}, \tag{5}$$

$$q_{evp,L} = k_L \frac{T - T_{sat}}{d_{int}}, \quad q_{evp,V} = k_V \frac{T - T_{sat}}{d_{int}}. \tag{6}$$

Here, the temperature gradient toward the interface was evaluated using the reconstructed interface. According to the method proposed by Kunkelmann et al.[18], the evaporative mass flux $\dot{m}_{evp}$ was distributed around the liquid-vapor interface, and the mass and energy source terms in Eqs. (1) and (3) were evaluated. These models were implemented using OpenFOAM 2.1.1 by Herbert et al.[20]. The governing equations were discretized using the finite volume method and implemented using the modified libraries of OpenFOAM.

The calculation domain was discretized using 80 grids in the radial direction and 2667 grids in the axial direction. The grid size in the radial direction was 1.25 μm and it was sufficient to resolve the liquid film.

**4. Definition of evaluated parameters**

4.1. Parameters describing bubble shape and liquid film

In Fig. 2, the definitions of the evaluation parameters for the bubble shape are shown. The bubble length $L_B$ is defined by the position of bubble head $z_{B,F}$ and bubble end $z_{B,E}$. The bubble head speed $U_B$ and expansion speed $U_{exp}$ are defined as

$$U_B = \frac{dz_{B,F}}{dt}, \tag{7}$$

$$U_{exp} = \frac{dL_B}{dt}. \tag{8}$$

Because of the low Reynolds number of the liquid phase, the bubble head speed and expansion speed are almost identical.

The liquid film was evaluated based on the interface shape $r_B(z)$ and defined as the flat area in the bubble. The flatness is defined using the interface shape as

$$\frac{dr_B}{dz} < 0.01. \tag{9}$$

To find the positions where Eq. (9) is satisfied, the edge position of liquid film on the front side $z_{L,F}$, on the end side $z_{F,E}$, and the length of liquid film $L_{LF} = |z_{L,F} - z_{L,E}|$ are defined.

The elongated bubble in the tube generally has a dimple at the backside after the liquid film. In this study, the region containing the dimple is called the rear edge region. The thickness of the liquid film was evaluated using the following three parameters:

$$\delta_{Full} = \frac{1}{L_B} \int_{z_{B,F}}^{z_{B,E}} \{R - r_B(z)\} dz, \tag{10}$$

$$\delta = \frac{1}{L_{LF}} \int_{z_{L,F}}^{z_{L,E}} \{R - r_B(z)\} dz, \tag{11}$$

$$\delta_0 = R - r_B(z_{L,F}). \tag{12}$$

Equation (10) denotes the average thickness of the liquid layer that surrounds the vapor bubble. This thickness can be determined regardless of whether liquid film exists. However, Eqs. (11) and (12) are used only after the liquid film is formed. $\delta$ denotes the average thickness of the liquid film, and $\delta_0$ denotes the thickness near the bubble head, which is often called the initial thickness. In this study, $\delta_0$ is called as the front edge thickness of the liquid film.

To evaluate the liquid film thickness, the capillary number is commonly used to express the ratio between the viscous force and surface tension. The capillary number is defined as

$$Ca = \frac{\mu U_B}{\sigma}. \tag{13}$$

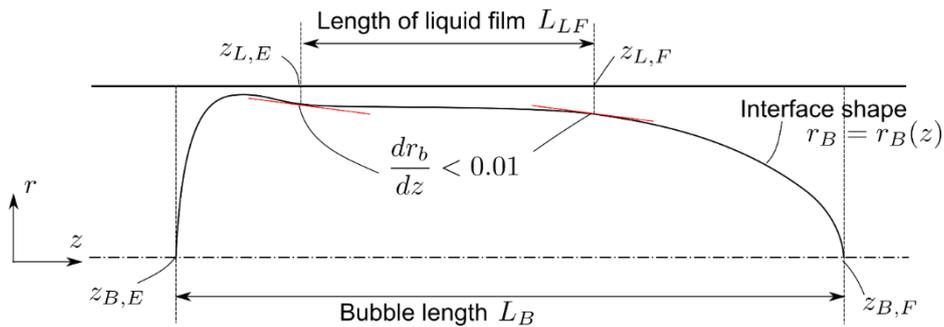

Fig. 2. Definition of parameters describing bubble shape

4.2. Parameters describing heat transfer via evaporating bubble

Fig. 3 shows the definition of the parameters describing the heat transfer via the evaporating vapor bubble. In this study, the wall and liquid phase were initially considered isothermal; only the vapor bubble was the source of heat transfer. The energy for phase change as latent heat can be evaluated using the variation in bubble volume $V_B$ and expressed as

$$Q_B = \rho_V h_{LV} \frac{dV_B}{dt}. \tag{14}$$

The effect of the evaporating vapor bubble on the heat transfer from the wall was analyzed in this study. The wall heat flux distribution $q_{wall}$ and heat transfer rate from the wall $Q_{wall}$ are defined as

$$q_{wall}(z) = -k_L \left. \frac{dT(z)}{dz} \right|_{r=R}, \tag{15}$$

$$Q_{wall} = 2\pi R \int q_{wall}(z) dz, \tag{16}$$

where the integral range of Eq. (16) is from the inlet to the outlet of the channel. The heat transfer rate from the wall can be decomposed into the heat transfer rate through the liquid film, rear edge, and wake, as shown in Fig. 3; it is expressed as

$$Q_{wall} = Q_{LF} + Q_{RE} + Q_{wake}. \tag{17}$$

Here, the boundary between the region of the liquid film and rear edge is defined as the end of the liquid film. The boundary between the region of the rear edge and wake is defined as the position where the liquid height is the same as that at the end of the liquid film. The average heat flux $q_{ave}$ through the liquid film was calculated using the following equation:

$$q_{ave} = \frac{1}{L_{LF}} \int_{z_{L,F}}^{z_{L,E}} q(z) dz. \tag{18}$$

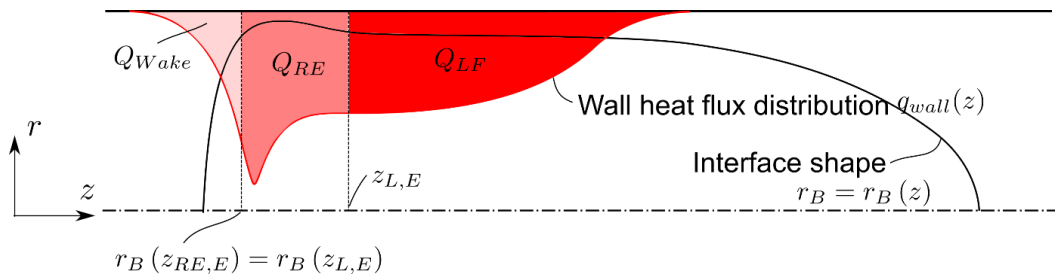

Fig. 3. Definition of parameters describing heat transfer component

### 4.3. Validation of definition of liquid film thickness

To validate the criterion for evaluating the liquid film given in Eq. (9), the calculated film thickness was compared with existing correlations. For this, a situation without evaporation was considered.

Bretherton's theory[21] and Taylor's law[22] were the existing correlations used for validation. In Bretherton's theory, the liquid film thickness was derived theoretically. This model is valid in regions with a small capillary number. Bretherton's theory is expressed as

$$\frac{\delta}{R} = 0.643(3Ca)^{\frac{2}{3}}. \tag{19}$$

Taylor's law is the extension of Bretherton's theory to express the liquid film in a high capillary number region was used, which is expressed as

$$\frac{\delta}{R} = \frac{0.643(3Ca)^{\frac{2}{3}}}{1+1.608(3Ca)^{\frac{2}{3}}}. \tag{20}$$

Fig. 4 shows the comparison of liquid film thicknesses obtained via numerical simulations and theoretical models. In this calculation, a long bubble was located in a channel with diameter of 200 μm, and a specific inlet velocity was set. As the two-phase working fluids, a mixture of silicone oil–air and liquid and vapors of FC-72 at 0.1013 MPa was selected. Table 1 lists the thermophysical properties of the working fluids used in this study. As shown in Fig. 4, the calculated thicknesses of the liquid film agrees with both Bretherton's theory and Taylor's law at low capillary numbers less than $10^{-2}$. At a high capillary number, the numerical results agreed with Taylor's law. Therefore, the numerical methods used and the evaluation criterion given in Eq. (9) were valid for calculating the liquid film thickness.

Table 1 Thermophysical properties of working fluids

| | | density [kg/m$^3$] | specific heat [J/(kg·K)] | thermal conductivity [W/(m·K)] | kinematic viscosity [m$^2$/s] | surface tension [N/m] | latent heat for vaporization [J/kg] |
|---|---|---|---|---|---|---|---|
| FC-72 | *liquid* | $1.62\times10^3$ | $1.10\times10^3$ | $5.22\times10^{-2}$ | $2.80\times10^{-7}$ | $8.27\times10^{-3}$ | $8.45\times10^4$ |
| (0.1013 MPa, 56.6°C) | *vapor* | $1.33\times10^1$ | $8.85\times10^2$ | $8.60\times10^{-3}$ | $7.08\times10^{-7}$ | | |
| Silicone oil | | $9.15\times10^2$ | - | - | $5.0\times10^{-6}$ | $1.97\times10^{-2}$ | - |
| Air | | 1.18 | - | - | $1.58\times10^{-5}$ | | - |

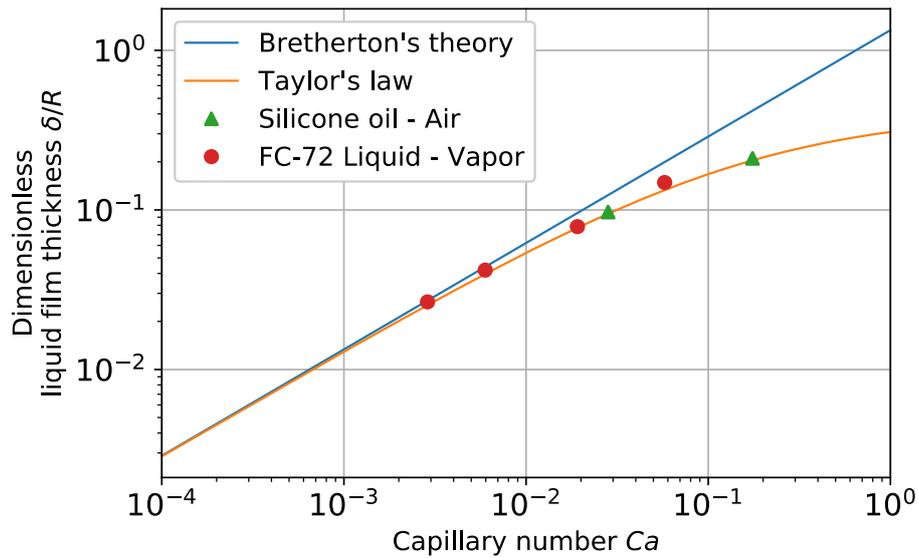

Fig. 4. Comparison of liquid film thickness calculated using numerical simulation and theoretical model

## 5. Results and discussion

This section presents the calculations conducted using the fluid of FC-72 at 0.1013 MPa as the representative dielectric liquid for electronics cooling.

5.1. Bubble dynamics and wall heat flux

Fig. 5 shows the time variation of bubble shape and heat flux distribution on the wall calculated by Eq. (15). In addition, the dots on the bubble shape line represent the edge of the liquid film. As shown Fig. 5, the heat flux on the wall was initially zero because the wall and liquid temperature was at same temperature. When the thermal boundary layer around a vapor bubble touched the wall, heat flux was observed on the wall. At 4.0 ms, the liquid film area was confirmed. The heat flux distribution had a peak value around the rear edge of the vapor bubble. The heat flux in the liquid film region was almost half the peak value. Additionally, as shown in the figure at 8.0 ms, no heat flux was observed around the bubble head. In this area, the thermal boundary layer around the vapor bubble did not touch the wall.

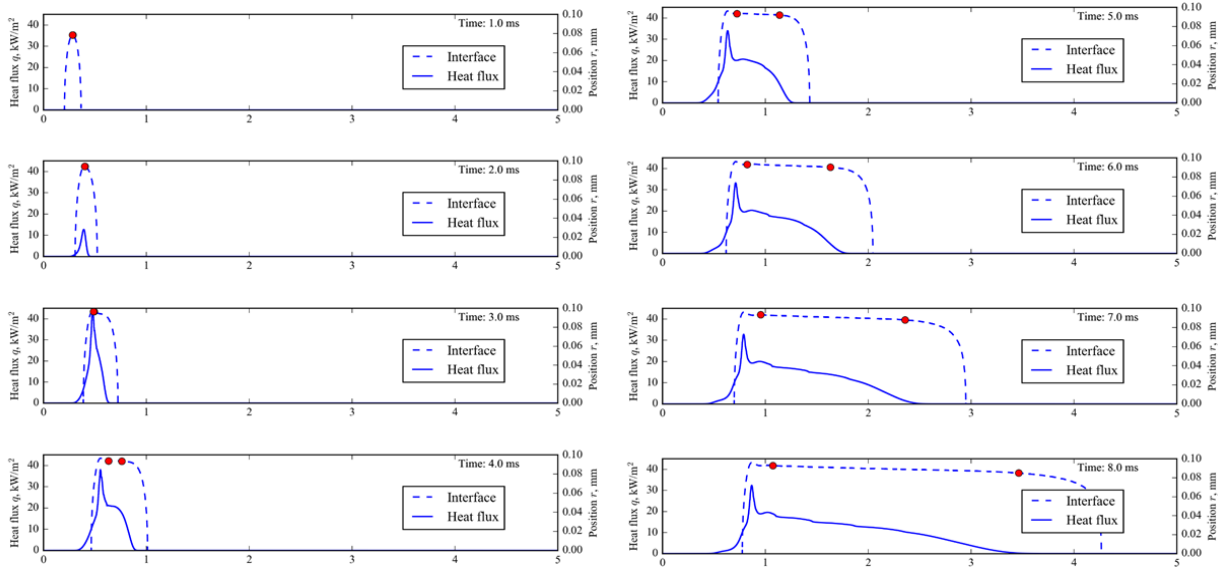

Fig. 5. Snapshots of bubble shape and distribution of wall heat flux in case of superheating at 2.5 K. The dots on the bubble shape represent the edge of the liquid film

Fig. 6 shows the time variation of bubble length under each superheat condition. As shown, the bubble length increased exponentially because the expansion rate depends on the surface area of a vapor bubble. Hence, the time scale of the expansion phenomena drastically differed under each superheat condition. Hereinafter, when comparing the time variation under each superheat condition, the bubble length is used instead of time.

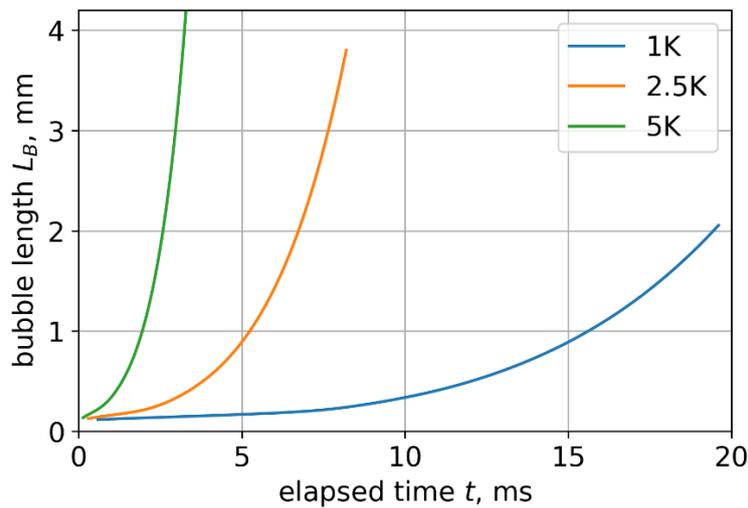

Fig. 6. Time variation of bubble length

5.2. Liquid film formation

Fig. 7 shows the time variation of the liquid film thickness at a superheat of 1 and 2.5 K. In the early stage of bubble expansion the average thickness of liquid layer $\delta_{Full}$ decreased because of the spherical expansion of the vapor bubble. Next, the confined growth of the vapor bubble and formation of the liquid film initiated at approximately 10 and 3 ms at 1 and 2.5 K, respectively. As shown in Fig. 7, both the front edge and average thickness of the liquid film increased. The higher the superheat, the larger is the difference between the initial and average thicknesses.

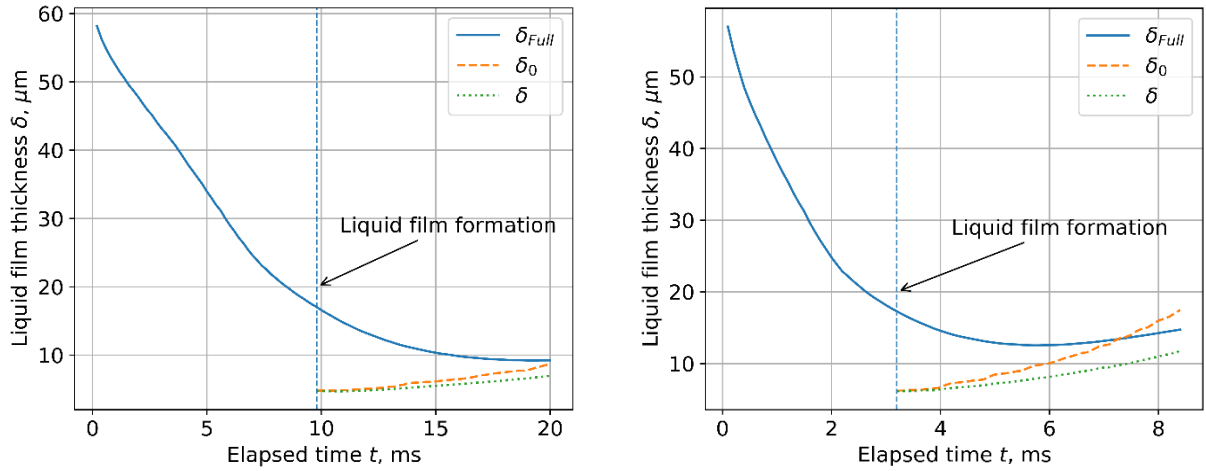

(a) superheat of 1 K  (b) superheat of 2.5 K

Fig. 7. Time variation of liquid film thickness

Fig. 8 shows the relationship between the average liquid film thickness and bubble length. As mentioned earlier (Fig. 6), the bubble length was used as an expression of the transient variation to normalize time durations. As shown in Fig. 8, the thicker liquid film was formed under a large superheat condition. According to Eqs. (19) and (20), a faster movement of the interface forms a thicker liquid film. The trend shown in Fig. 8 indicates that the evaporation-based thinning effect on the liquid film thickness is relatively small during the formation of the liquid film. Additionally, this result implies that the heat transfer coefficient decreases with increasing superheat because the heat transfer coefficient is roughly expressed as $k_L/\delta$.

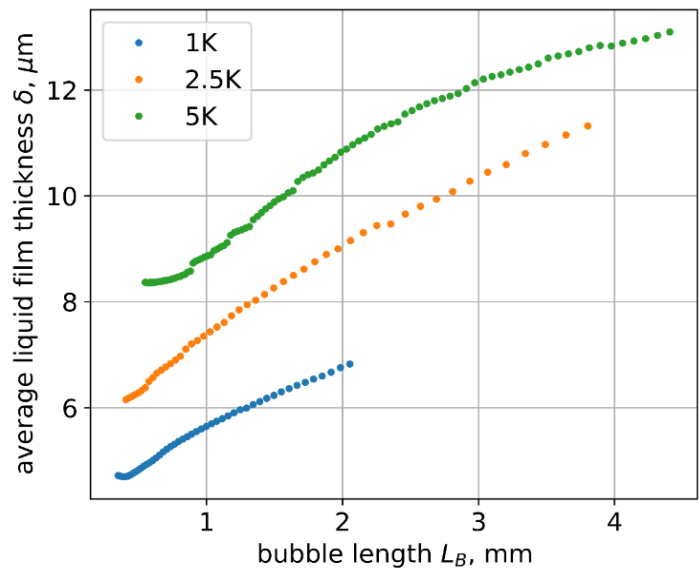

Fig. 8. Relationship between the average liquid film thickness and bubble length

Fig. 9 shows the relationship between bubble and liquid film lengths. This relation describes the fraction of the liquid film length to the whole bubble length. As shown, the fraction of the liquid film length increased with low superheat. This trend can be confirmed in Fig. 10, which shows the heat flux distribution and interface shape for a 2-mm bubble length. The lower the superheat condition, the flatter is the liquid film. Therefore, at a larger expansion speed, the bubble head became sharper and the fraction of the liquid film area decreased.

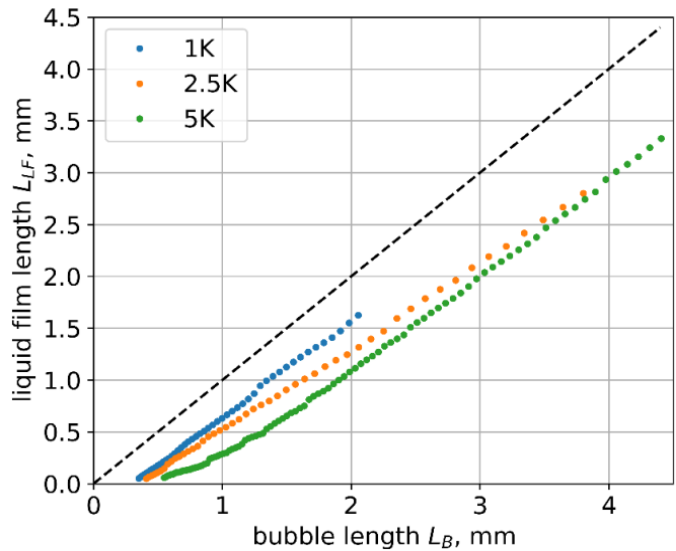

Fig. 9. Relationship between the bubble and liquid film lengths

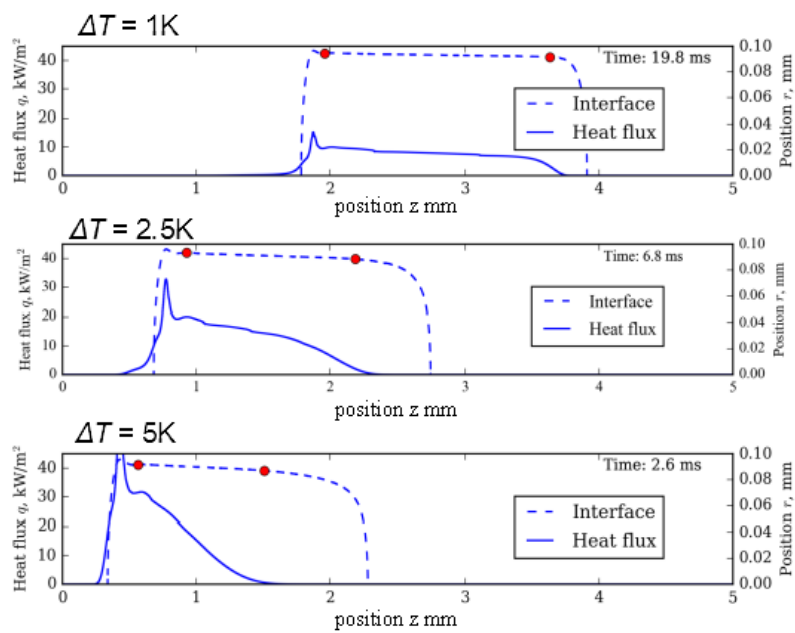

Fig. 10. Numerical results of heat flux distribution and interface shape for a 2-mm bubble length

Fig. 11 shows the relationship between the capillary number and dimensionless thickness of the liquid film. The dimensionless thickness was derived by scaling the thickness according to the channel radius, and the calculated values were compared with those of previous studies: Bretherton's theory [21], Taylor's law [22], and Han's correlation [4]. As shown in Fig. 11(a), the front edge thickness in the present study agreed with Taylor's law in the case of superheat of 1 and 2.5 K. As the superheat increased, the front edge thickness decreased more than that defined by Taylor's law as a larger acceleration affects the liquid film thickness. As shown in Fig. 11(b), the average thickness was lower than the existing correlations in the region with a high capillary number. This is caused by the variation in velocity during the expansion process. These results indicated that the spatial distribution of the liquid film thickness becomes important to derive the representative value of liquid film thickness and heat flux. Additionally, the average thickness seems to have a better correlation with the capillary number than the front edge thickness. This result implies that the average thickness is predictable by using the capillary number.

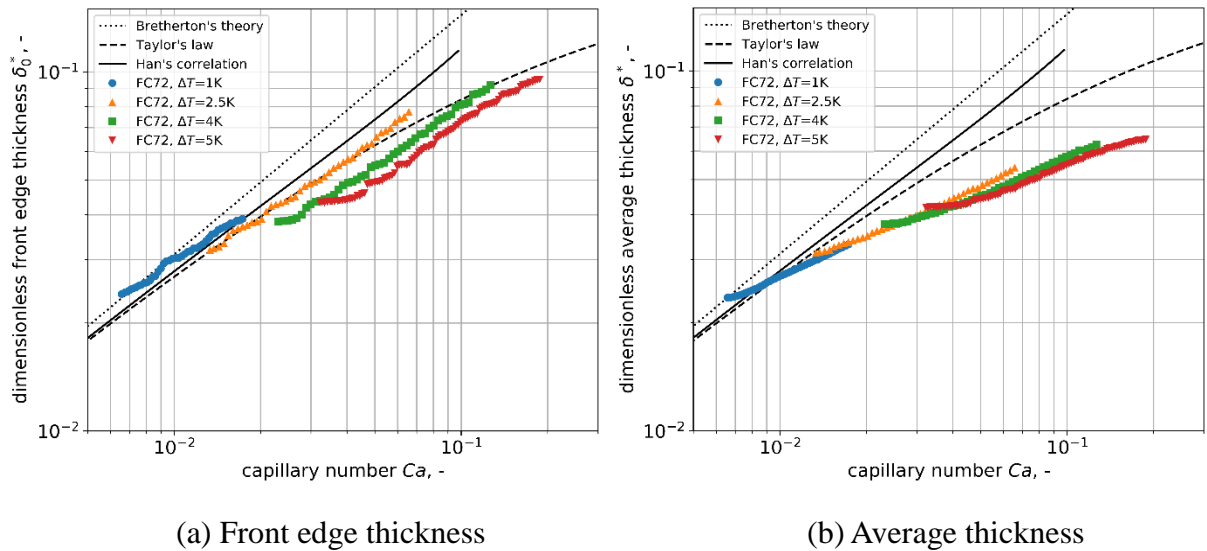

(a) Front edge thickness  (b) Average thickness

Fig. 11. Comparison of the liquid-film thickness derived in numerical simulation with that derived through theoretical and empirical correlations

5.3. Heat transfer

By analyzing the transient heat flux distribution, the heat transfer components in the evaporating process is discussed. Fig. 12 shows the time variation of each component of heat transfer rate under the conditions of superheat of 1 and 2.5 K. The definition of each heat transfer rate is expressed in Eq. (17). As shown in Fig. 12, in both the superheat conditions, the heat transfer rates at the rear edge $Q_{RE}$ and wake $Q_{wake}$ were kept constant. Therefore, the variation in the heat transfer rate from the wall is caused by the increments of heat transfer rate through the liquid film.

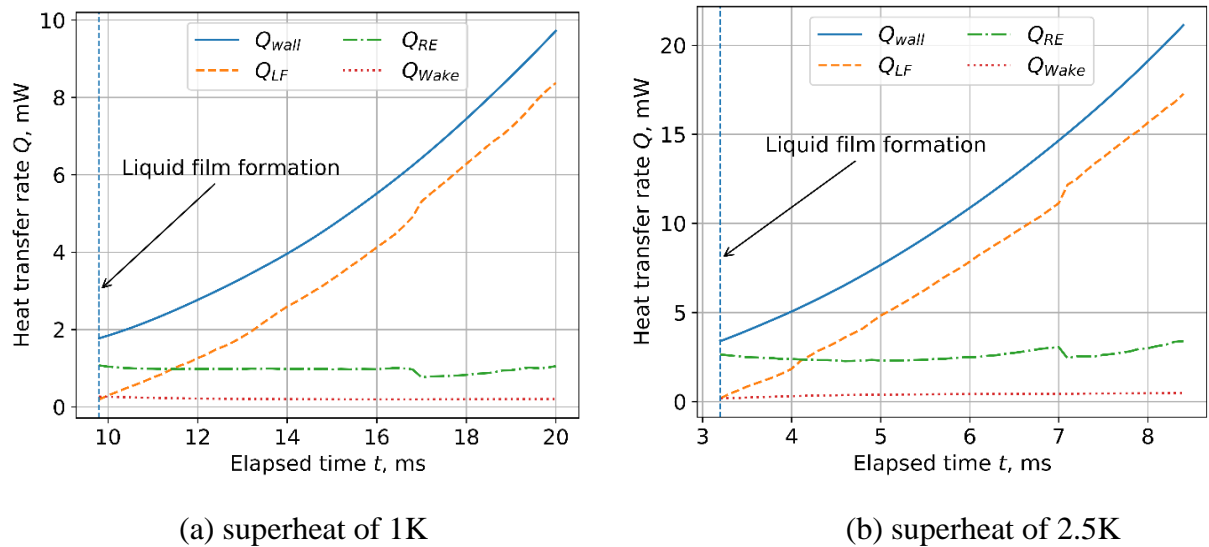

(a) superheat of 1K  (b) superheat of 2.5K

Fig. 12. Time variation of each component of heat transfer rate

Fig. 13 shows the time variation of energy for phase change as latent heat and heat transfer rate from the wall and through liquid film under the superheat conditions of 1 and 2.5 K. The energy for phase change as latent heat was calculated using Eq. (14). As shown in Fig. 13(a) and (b), the difference between $Q_B$ and $Q_{wall}$ increased with time. Especially, in the case of 2.5-K superheat, the difference between $Q_B$ and $Q_{wall}$ was larger than that in the case of 1-K superheat. This fact implies that the contribution of heat transfer rate from the wall was reduced in higher superheat conditions.

To evaluate the contribution of heat transfer rate from the wall to bubble expansion, energy ratio $Q_{wall}/Q_B$ was compared among several superheat cases. Fig. 14 shows the relationship between energy ratio of heat transfer rate from wall to latent heat and bubble length. Here, the bubble length was used to express the time variation. As shown in the figure, in the case of superheat of 1 K, the energy ratio was kept at 0.9. This implies that the vapor supply for bubble expansion is almost provided through the heat transfer from the wall; namely, the heat transfer due to liquid film evaporation contributes toward the bubble expansion. Moreover, the energy ratio decreased with increasing superheat. Especially, when superheat was 5 K, the contribution of the wall heat transfer was reduced to less than 30%. Therefore, the evaporation-based bubble expansion did not contribute much toward the heat transfer from the heated wall.

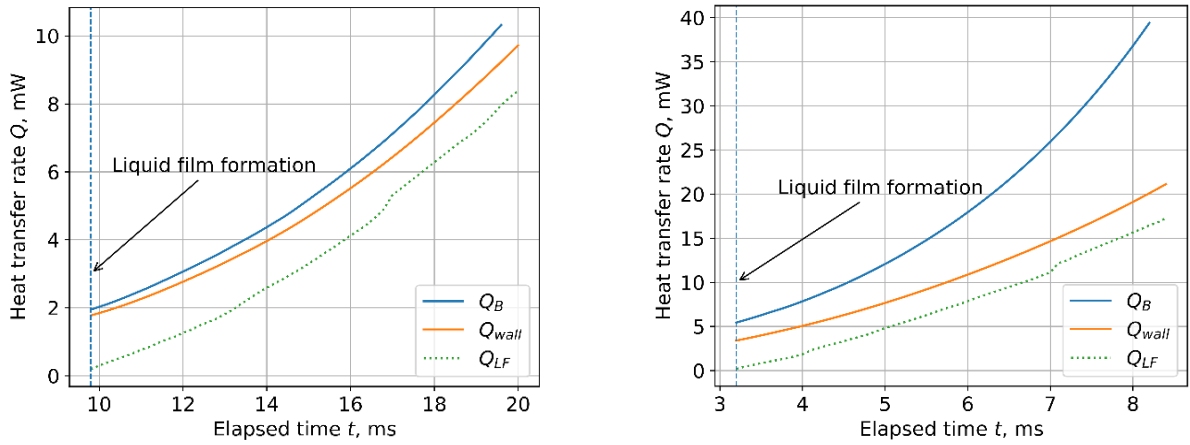

(a) superheat of 1 K            (b) superheat of 2.5 K

Fig. 13. Time variation of energy for phase change as latent heat and heat-transfer rate from wall and through liquid film

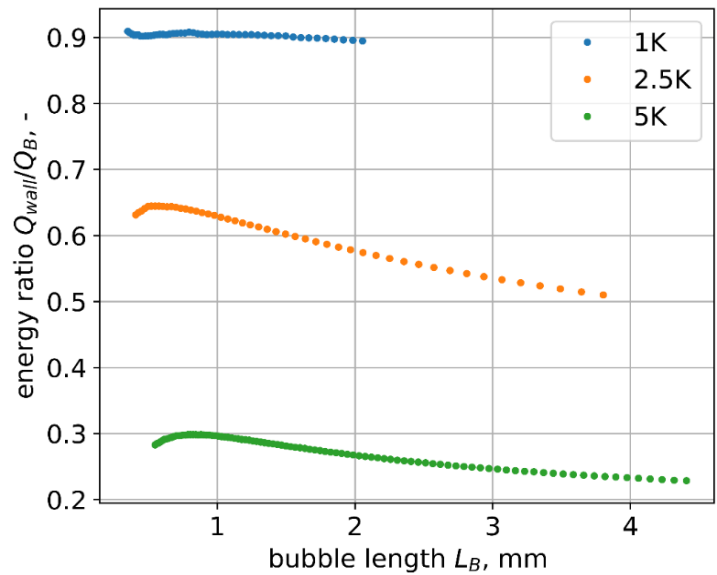

Fig. 14. Relationship between energy ratio of heat transfer rate from wall to latent heat and bubble length

In general treatment, the heat flux or heat transfer coefficient is calculated based on the liquid film thickness [6, 23]. In this treatment, the linear temperature distribution in the liquid film is assumed. To validate this assumption, in this study, the heat flux was derived using the following two methods, and the values were then compared. First, the heat flux was calculated based on the distribution of liquid film thickness, expressed as

$$\bar{q}_\delta = \frac{1}{L_{LF}} \int k_L \frac{\Delta T}{\delta(z)} dz. \qquad (21)$$

Second, the heat flux was directly calculated through temperature distribution, as shown in Eq. (18). Fig. 15 shows the relationship between average heat flux based on liquid film and bubble lengths. The averaged heat flux was defined as the averaged value of heat flux distributed in the liquid film area. As shown in this figure, the average heat flux decreased with increase in the bubble length, which was caused by the increase of the average liquid film thickness during bubble expansion. Additionally, the averaged heat flux was compared with the heat flux derived from liquid film thickness calculated using Eq. (21). As shown in Fig. 15, when the superheat was 1 K, the two heat fluxes correspond to each other, indicating that the heat flux can be predicted from the liquid film thickness. In contrast, in the cases of superheats of 2.5 and 5 K, large differences can be observed between the two heat fluxes. These results indicate that the assumption of linear temperature distribution in the liquid film is not yet valid. Magnini et al.[24] modeled the variation in the thermal boundary layer formed near the wall heated by a constant heat flux by using the transient heat conduction equation. Their work also indicated that the assumption of a one-dimensional steady-state temperature distribution of the liquid film led to the overestimation of the heat transfer coefficient. As shown in Fig. 10, especially in the case of the superheat of 5 K, the heat flux around the front edge of the liquid film was smaller than that around the backside of the liquid film. Near the bubble head, the thermal boundary layer around the bubble became thinner owing to the rapid expansion of the vapor bubble. No evaporative heat transfer occurred in the liquid film if the thermal boundary layer was thinner than the liquid film. These results indicated that there is a region where heat transfer between heated wall and liquid-vapor interface does not occur even in the liquid film defined geometrically. Additionally, when the vapor bubble expanded in high superheated liquid, the rapid growth thickened the liquid film, which prevented the heat transfer between the liquid-vapor interface and heated wall.

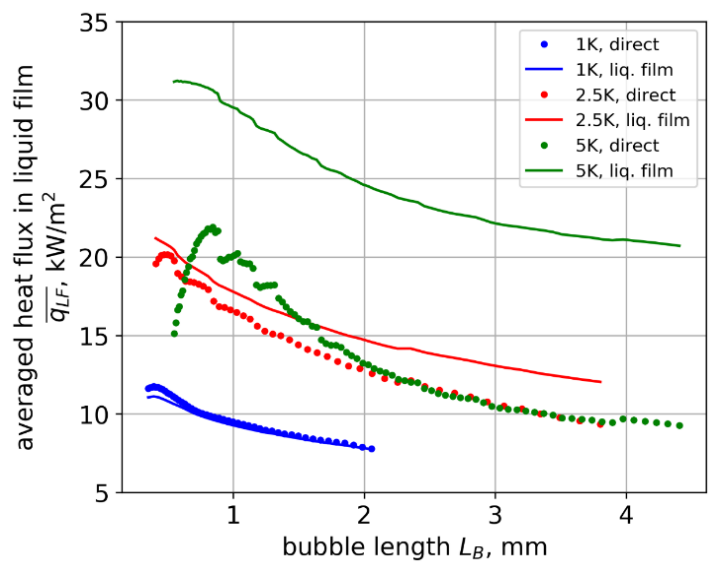

Fig. 15. Comparison between two types of averaged heat flux through liquid film: one was calculated directly through numerical simulation while the other was calculated from the distribution of the liquid film thickness

## 6. Conclusions

In this study, the formation process of a liquid film and its heat transfer during vapor bubble expansion was investigated through numerical simulations. The working fluid was assumed as a typical electrical insulated fluid of FC-72 at 0.1013 MPa. The expansion process of the vapor bubble in the uniformly superheated liquid was simulated. The obtained results were as follows:

- Owing to the exponential increment of the expansion velocity, the averaged liquid film thickened with time. In the higher superheating cases, the expansion speed was larger and the liquid film formed was thicker. Owing to a large expansion speed, the ratio of liquid film length to bubble length decreased because the bubble head sharpened.
- The front edge thickness in the present study agreed with the Taylor's law in the case of superheating at 1 and 2.5 K. As the superheating increased, the front edge thickness became thinner than that derived by Taylor's law because a larger acceleration affects the liquid film thickness. The average thickness showed better correlation with the capillary number than the front edge thickness, implying that the average thickness is predictable by using a capillary number.
- To evaluate the contribution of heat transfer rate from the wall toward bubble expansion, the energy ratio of heat transfer rate from the wall to latent heat consumption due to phase change was calculated. In the case of superheating at 1 K, the energy ratio was kept at 90%, indicating that the vapor for the bubble expansion was supplied by the heat transfer from the wall. Namely, the heat transfer through liquid film evaporation contributes toward bubble expansion. Moreover, the energy ratio decreased with increase in the superheat value and the contribution of wall heat transfer was reduced to less than 30%. Therefore, the evaporation-based bubble expansion did not contribute much toward the heat transfer from the heated wall.
- Furthermore, the assumption of the linear temperature distribution in the liquid film was validated. When the superheat was 1 K, the heat flux through the liquid film could be predicted with respect to the liquid film thickness. However, in the cases of superheating at 2.5 and 5 K, the assumption of linear temperature distribution in the liquid film was not valid. These results indicated that there is a region where the heat transfer between the heated wall and liquid−vapor interface does not occur even in the liquid film is defined geometrically.

In general, the relationship between the thickness of the thermal boundary layer of the bubble and liquid film thickness should be examined to predict the cooling effect of this phenomenon. When the vapor bubble expands in a highly superheated liquid, its rapid growth thickens the liquid film, which prevents the heat transfer between the liquid–vapor interface and heated wall. The characteristics of the liquid film and evaporative heat transfer of the expanding bubble by evaporation should be generalized by the method proposed in this paper.

*Nomenclatures*

| | | |
|---|---|---|
| $Ca$ | capillary number | - |
| $c$ | specific heat | J/(kg·K) |
| $c_F$ | compression parameter | - |
| $d$ | distance | m |
| $F$ | volume fraction of liquid | - |
| $h_{LV}$ | latent heat | J/kg |
| $k$ | thermal conductivity | W/(m·K) |
| $L$ | length | m |
| $\dot{m}$ | mass flux | kg/(m²·s) |
| $p$ | pressure | Pa |
| $Q$ | heat transfer rate | W |
| $q$ | heat flux | W/m² |
| $R$ | radius | m |
| $r_B$ | interface shape | m |
| $U$ | speed | m/s |
| $\boldsymbol{u}$ | velocity vector | m/s |
| $T$ | temperature | K |
| $t$ | time | s |
| $V$ | volume | m³ |
| $z$ | z coordinate | m |

Greek

| | | |
|---|---|---|
| $\delta$ | liquid film thickness | m |
| $\kappa$ | curvature | 1/m |
| $\mu$ | viscosity | Pa·s |
| $\rho$ | density | kg/m³ |
| $\Sigma_e$ | energy source term | W/m³ |
| $\Sigma_V$ | volume source term | 1/s |
| $\sigma$ | surface tension | N/m |

Subscript

| | |
|---|---|
| *ave* | average |
| *B* | bubble |
| *E* | end |
| *evp* | evaporation |
| *exp* | expansion |
| *F* | front |
| *int* | interface |
| *L* | liquid |
| *LF* | liquid film |
| *RE* | rear edge |
| *sat* | saturation |
| *V* | vapor |
| *wake* | wake |
| *wall* | wall |

*Acknowledgments*

This work was supported by JSPS KAKENHI Grant Numbers JP25820054 and JP17K06183 and Young Researcher Overseas Visits Program for Vitalizing Brain Circulation R2503.